# Wide range and tunable linear TMR sensor using two exchange pinned electrodes

B. Negulescu[a], D. Lacour[a], F. Montaigne[a], A. Gerken[b], J. Paul[b], V. Spetter[b], J. Marien[b], C. Duret[c], M. Hehn[a]

*[a] Institut Jean Lamour, CNRS - Nancy-Université - UPV-Metz, Boulevard des Aiguillettes BP 70239, F-54506 Vandoeuvre lès Nancy, France*

*[b] Sensitec GmbH, Hechtsheimer Str. 2, 55131 Mainz, Germany*

*[c] SNR Roulements, BP 2017, FR-74010, Annecy, France*

A magnetic tunnel junction sensor is proposed, with both the detection and the reference layers pinned by IrMn. Using the differences in the blocking temperatures of the IrMn films with different thicknesses, crossed anisotropies can be induced between the detection and the reference electrodes. The pinning of the sensing electrode ensures a linear and reversible output. It also allows tuning both the sensitivity and the linear range of the sensor. The authors show that the sensitivity varies linearly with the ferromagnetic thickness of the detection electrode. It is demonstrated that an increased thickness leads to a rise of sensitivity and a reduction of the operating range.

**Keywords:** magnetic field sensor, TMR, AlOx tunnel junction, sensitivity, linearity range, IrMn, exchange bias, blocking temperature

**PACS numbers:** 73.43.Jn ; 75.47.m ; 75.50.Ee ; 75.60.Nt

Magnetic field sensors are used in a wide range of applications such as read heads in the magnetic data storage industry, as position, speed and rotation angle detectors in the automotive industry, as electric current detectors or as magnetic flux leakage detectors in the non-destructive testing technology. [1-3] Nowadays, the majority of the magnetic sensors developed for industrial applications are based on the Hall effect, the anisotropic or the giant magnetoresistance effects, due to their ease of integration and low production costs. However, the sensitivity can be largely increased by using magnetic tunnel junctions as field detectors. Also, the relatively high resistance of the tunnel magnetoresistance (TMR) devices ensures a low power consumption of the sensor which constitutes a major advantage for mobile applications and green electronics. Besides their use in read heads of hard disk drives, currently the TMR sensors are mainly developed for biological applications, [4] where high sensitivity and low noise are essential characteristics.

The key feature of such sensors is the linear and nonhysteretic electrical response to an external magnetic field. In the case of a TMR device, the linear characteristic originates from a perpendicular configuration of the electrodes magnetizations at zero applied field, whereas the nonhysteretic behaviour results from the rotation of the detection electrode's magnetization. In order to control the reversal mechanism of the detection layer, our group has previously proposed the use of an exchange coupled ferromagnetic layer.[5] A reversible and linear magnetoresistive response was demonstrated in a device where a hard magnetic CoPt film was used as reference electrode. However, in this initial demonstrator, the two electrodes were switching independently only in a small field range, leading to a reduced operating range.

In this letter, we show a fully functional TMR sensor based on a multilayer stack including two IrMn antiferromagnetic (AF) thin films. The sensitivity and the linear range are easily tunable by only varying the thickness of the pinned ferromagnetic detection layer.

The pinning directions of the two AF layers have to be set orthogonal in order to achieve a linear response. If the bottom and the top exchange coupled interfaces have different temperature stabilities, this configuration can be obtained through two successive annealing steps at different temperatures with crossed applied fields. The blocking temperatures ($T_B$) of the IrMn layers are known to be dependent on the AF's layer thickness as well as on the top or bottom pinned configurations [6]. They rise with increased AF thickness and are higher in the bottom configuration compared to the top one. [7] In the mean time and in the studied range of IrMn thicknesses (above 5nm), the exchange bias field ($H_{ex}$) decreases as a function of the IrMn thickness, meaning that a high AF thickness reduces the pinning strength.

Aiming at high temperature stability of the reference electrode, its magnetization is pinned by a thick IrMn layer (15 nm) while a thinner IrMn layer (6 nm) is part of the detection electrode. In order to compensate the decrease of $H_{ex}$ due to the increased thickness of the AF, the reference electrode is placed at the bottom side of the stack [8] and a synthetic antiferromagnetic (SAF) structure is used instead of a simple ferromagnetic layer. The SAF also gives the advantages of reducing the blocking temperature dispersion of the reference electrode,[9] of increasing its rigidity against rotation [10] and of decreasing its stray field. These properties ensure the magnetization stability of the reference electrode and the decoupling between the two electrodes.

The detection electrode consists of a top pinned CoFe x nm / IrMn 6nm bilayer while the reference electrode structure is: IrMn 15nm / CoFe 2.5nm / Ru 0.8nm / CoFe 3nm. The samples are deposited on 5 inch wafers in a DC magnetron sputtering system with a base pressure of $2\times10^{-8}$ Torr and typical deposition pressures of 2-3 mTorr at the Sensitec research division. An $Al_2O_3$ tunneling barrier is obtained by reactive RF oxidation of a 2 nm Al layer at a power of 50 W. The samples are annealed post deposition at chosen temperature (up to 265 °C) and magnetic field values in an $N_2$ atmosphere oven. The magnetic tunnel junction are then prepared in circular shapes with diameters varying from 40 μm to 100 μm using conventional photolithography and ion mill processes. After optimizing the oxidation conditions, a maximum TMR of 40 ± 3 % and a RxA product of 22 ± 6 MΩ μm² are measured. Considering the thickness of the $Al_2O_3$ layer in our samples (2nm measured by transmission electron microscopy) and the dependence of the TMR values upon the barrier thickness and height,[11] this value is state of the art.[12]

After annealing the samples at 265 °C under 10 kOe the exchange coupling effect is activated and both electrodes magnetizations are pinned parallel to the applied field direction. In this magnetic state, the m(H) loops were measured at different temperatures (starting from room temperature up to 300 °C) in order to calculate the variation of the exchange coupling energy ($J_{ex}$) vs. temperature (presented in FIG.1). These measurements allow determining the blocking temperature defined as the temperature at which $J_{ex}$ becomes equal to zero. $J_{ex}$ of the detection bilayer is calculated from the measured exchange bias field ($H_{ex}^{detection}$) as: $J_{ex} = M_S t_{CoFe} H_{ex}^{detection}$, where $M_S$ and $t_{CoFe}$ are the saturation magnetization and the thickness of the detection CoFe film respectively. In the case of the reference electrode, $J_{ex}$ is obtained by fitting the measured m(H) curves using a macro spin model of the pinned SAF.[13] From figure 1, the $T_B$ value of the detection electrode is about 200 °C, while the reference electrode is pinned up to (275 ± 15) °C. Consequently the unidirectional anisotropies induced by the AF layers in the two ferromagnetic electrodes can be independently tuned by properly choosing the annealing temperatures. A first annealing at high temperature initializes both layers in the same direction. A second annealing at a temperature between 200 °C and 275 °C and with the field applied perpendicular to the first annealing field direction is done afterwards, in order to define an orthogonal pinning for the detection electrode's magnetization. The TMR curves measured on the same wafer after a first annealing at 265

°C (-●-) and a second annealing at 240 °C (-○-) are shown in FIG. 2. In the low field range, the variation of resistance is only due to the magnetization reversal of the top sensing layer. The shift of the curve in negative fields observed after the first annealing procedure corresponds to the exchange coupling field for the detection CoFe 5nm / IrMn 6nm bilayer: $H_{ex}^{detection}$ = 270 Oe. After the second annealing procedure, the TMR variation is measured with the field applied perpendicular to the direction of the detection electrode exchange anisotropy. The detection electrode's magnetization rotates in field, resulting in the targeted linear variation of the resistance required for field sensing applications. Similar TMR curves are measured in the high field range, signifying that the anisotropy of the reference electrode is not affected by the second annealing step.

The total sensitivity of a TMR sensor with pinned sensing layer [5] is inversely proportional to $H_{ex}^{detection}$ when $H << H_{ex}^{detection}$. The sensitivity can be written as:

$$S = \frac{\Delta R}{\Delta H} = \frac{R_P - R_{AP}}{2} \times \frac{1}{H_{ex}^{detection}} = \frac{R_P - R_{AP}}{2} \times \frac{M_S t_{CoFe}}{J_{ex}},$$

with $R_P$ and $R_{AP}$, the resistances of the tunnel junction with parallel and respectively antiparallel alignments of the electrodes magnetizations enclosing the tunnel barrier.

This variation law was experimentally verified on a set of samples with 3, 5, 7 and 9 nm thick top CoFe layers. After the first annealing procedure, all MTJs show TMR values of 40 % with the exception of the structure with the thinnest top CoFe layer, where the TMR attains only 27 %. The decrease of the TMR value for devices with such thin electrodes can be related to the loss of spin polarization, as already observed for structures with free sensing layers thinner than 2-3 nm. [14] In these devices the TMR reduction was associated with a decrease of the magnetization in the saturation state, but in our samples the $M_s$ values are constant for CoFe layers with thicknesses larger than 2 nm. This rules out the above mentioned explanation for the TMR decrease. Oxygen assisted Mn diffusion towards the barrier during the thermal annealing procedure could also induce TMR losses. [15] This diffusion mechanism is predominant in the top electrode that is generally not as well textured as the bottom one. [7, 15]

After reorienting the pinning direction of the detection electrode perpendicularly to the reference electrode anisotropy, all the samples show a linear resistance variation (as presented in FIG. 3). The sensitivity varies as a function of the detection CoFe thickness. For the thickest layer, a 10 % TMR linear variation is obtained in a range of ± 40 Oe. The sample with only 3 nm top CoFe has a much larger linearity range, extending between ± 220 Oe. For this sample, the TMR variation on the linear range is 4 %.

The hysteresis evolution with the external field range is presented in the inset of FIG. 3. In fields smaller than 400 Oe, the hysteresis increases linearly with the measuring range. This is the regime where the magnetization of the reference layer is stable and only the detection layer rotates in field. For higher fields, the hysteresis variation slows down and attains a maximum value of 10 Oe corresponding to a complete reversal between the parallel and the antiparallel states of the entire structure.

In FIG. 4 both the linearity range [16] and the sensitivity are plotted as a function of the CoFe detection layer's thickness. As theoretically predicted, we observe a linear variation of $S$ upon thickness. The operating range increases with the decrease of the CoFe thickness, as implied by the theoretical condition for linearity: $H << H_{ex}^{detection} = J_{ex}/(M_S t_{CoFe})$. Since the SAF structure used in the sensor has a stable antiparallel configuration between -2000 Oe and +500 Oe, the linearity range is not limited by the reference electrode. It is restricted only by the theoretical response of the detection layer, non linear for high fields.

In conclusion, we show a fully functional TMR sensor based on a multilayer stack including two antiferromagnetic (AF) thin films. Using the difference in the blocking temperatures of IrMn layers with different thicknesses, crossed pinning directions can be induced in the top and the bottom electrode of a MTJ device. The linear TMR (H) variation can be tuned by changing the thickness of the sensing ferromagnetic layer. The sensitivity is changed over a decade and the linear field range by a factor of 5 following the theoretical predictions. This sensor architecture appears to be fully tuneable within a wide field range. Furthermore, since the magnetic response of the sensing layer is not driven by the shape of the device, this sensor is fully scalable from millimeter down to nanometer.

This work was financed by the ANR CAMEL project. The authors would like to thank G. Lengaigne for help with the experiments.

[16] The linearity range is determined as following: first the TMR (H) curves are fitted with a linear function in different field ranges; for each fit, the non linearity is computed as the maximum deviation from the linear fit divided by the maximum TMR value in the considered field window, expressed in percentage; finally the linearity range is defined by a non linearity smaller than 5%.

**Figures caption :**

FIG. 1: Variation of the exchange coupling constant as a function of temperature for the top detection electrode (filled symbols) and the bottom reference electrode (empty symbols) of a substrate /Ta 5 / Ru 2 / IrMn 15 / CoFe 2.5 / Ru 0.8 / CoFe 3 / AlOx 2 / CoFe 5 / IrMn 6/ Ru 4 MTJ structure (the thicknesses are expressed in nm).

FIG. 2. TMR curves measured in the easy axis direction of the reference electrode after the first anneal and after the second anneal.

FIG.3. RxA variations of 4 MTJs with different thicknesses of the detection CoFe layers in the ± 50 Oe field range. The inset shows the coercivity variation as a function of the measuring field range for a sensor with a linear output between -100 Oe and + 100 Oe (5nm thickness of detecting CoFe layer).

FIG. 4. Sensitivity (top part) and linearity range (bottom part) vs. CoFe detection layer's thickness. The lines are fits with a linear law for the top part and a $1/t_{CoFe}$ law for the bottom one.

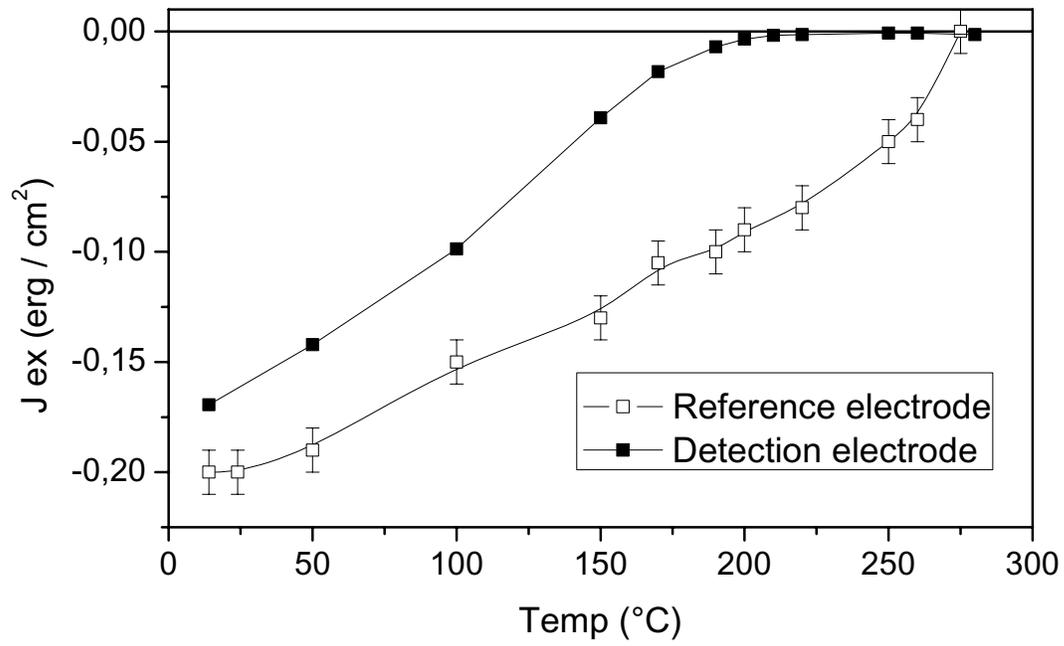

Fig. 1

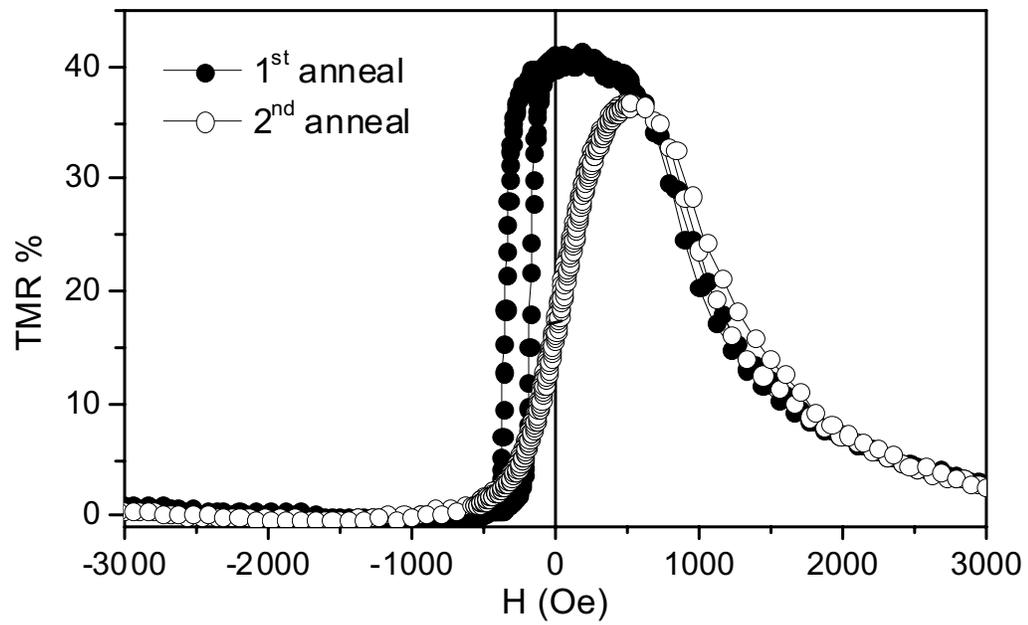

Fig. 2

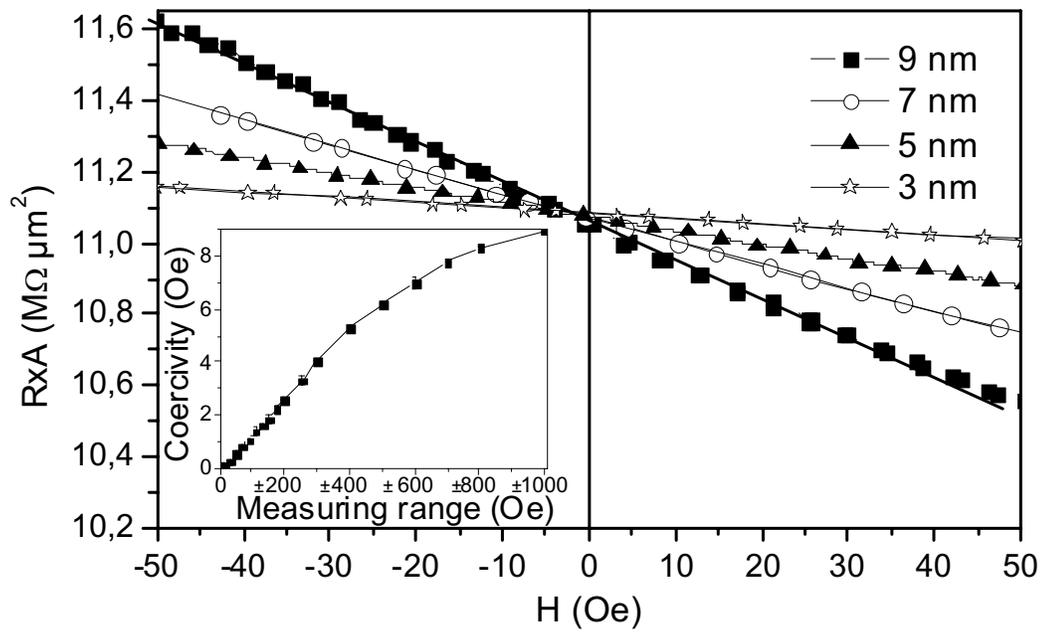

Fig. 3

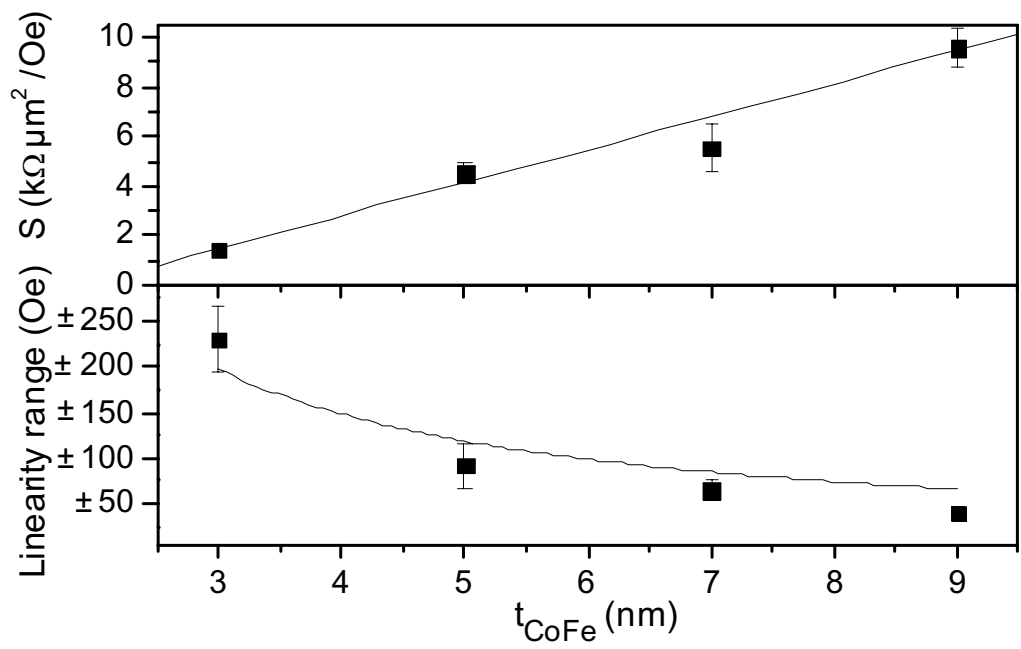

Fig. 4